\documentclass[aps,pra,twocolumn,superscriptaddress]{revtex4-1}

\usepackage[pdftex]{hyperref}
\usepackage{graphicx}
\usepackage{amsmath,amssymb,amsfonts}
\usepackage{units}
\usepackage{bm}
\usepackage{bbm}
\usepackage{xspace}

\newcommand{\PT}{\ensuremath{\mathcal{PT}}\xspace}
\newcommand{\lM}{\ensuremath{\lambda_M}\xspace}
\newcommand{\lMp}{\ensuremath{\lambda_M'}\xspace}
\newcommand{\lMh}{\ensuremath{\hat{\lambda}_M}\xspace}
\newcommand{\mul}{\ensuremath{\mu_l}\xspace}
\newcommand{\mulp}{\ensuremath{\mu_l'}\xspace}

\begin{document}

\title{Hybrid Breathers in Nonlinear \PT-Symmetric Metamaterials}

\author{Sascha B\"ohrkircher}
\email{sascha.boehrkircher@itp1.uni-stuttgart.de}
\affiliation{Institut f\"ur Theoretische Physik 1, Universit\"at Stuttgart,
  70550 Stuttgart, Germany}
\author{Sebastian Erfort}
\affiliation{Institut f\"ur Theoretische Physik 1, Universit\"at Stuttgart,
  70550 Stuttgart, Germany}
\author{Holger Cartarius}
\affiliation{Institut f\"ur Theoretische Physik 1, Universit\"at Stuttgart,
  70550 Stuttgart, Germany}
\affiliation{Physik und ihre Didaktik, 5.\ Physikalisches Institut,
  Universit\"at Stuttgart, 70550 Stuttgart, Germany}
\author{G\"unter Wunner}
\affiliation{Institut f\"ur Theoretische Physik 1, Universit\"at Stuttgart,
  70550 Stuttgart, Germany}

\date{\today}

\begin{abstract}
  On a two-dimensional planar parity-time-(\PT-)symmetric nonlinear magnetic
  metamaterial, consisting of split-ring dimers with balanced gain and loss,
  discrete breather solutions can be found. We extend these studies and by
  numerical calculations reveal the existence of further stable, long-lived
  oscillations, with certain frequencies, in the breather spectrum. We describe
  these oscillations in terms of an analytical breather theory, and show that
  they can be interpreted as superpositions of a breather oscillation and a
  plane wave. We coin the term  `hybrid breather' solutions for these 
  solutions.
\end{abstract}

\maketitle

\section{Introduction}
Metamaterials are artificially created lattice structures which can exhibit
non-intuitive properties, such as negative refractive indices or
unidirectional transparency \cite{Atmd,Fowdm,Shelby77,Shalaev2007,Lin2011,%
  Schurig2006}. They can be described by multiple electronic circuits, the
so-called split-ring resonators which are coupled electro-magnetically to
each other \cite{Sydoruk2006,Hesmer2007,Sersic2009}. The properties of these
metamaterials can be tuned by varying the dielectric material in the different
split-ring resonators of the lattice or spacings between the resonators
\cite{Tsironis2006}.
Due to the possibility of custom tailoring the basic properties of
metamaterials, they can even be used to realize discrete
parity-time-(\PT)-symmetric nonlinear metamaterials with balanced gain and
loss \cite{Schindler2011,Tsironis2013}.

The concept of \PT symmetry was introduced two decades ago \cite{Bender1998}
and ever since has attracted increasing attention, both theoretically
and experimentally (see, e.g., the recent book by Bender \cite{Bender2018} and
references therein).  $\mathcal PT$ symmetry has been realized
experimentally, e.g., in optics \cite{Ruter2010}. 

Every material unavoidably has loss. However, with a compensating equal
amount of gain stationary behavior still can be maintained, and thus
balanced gain and loss is a very common possibility to realize $\PT$
symmetry.

In  systems governed by nonlinear wave  equations highly localized
and stable oscillations, so-called breather solutions, have been found.
Examples are the Akhmediev breather as a solution of the Gross-Piteavskii
equation \cite{Akhmediev1987}, the Kuznetsov-Ma breather as a solution of the
Sine-Gordon equation \cite{Kuznetsov1977}, or discrete breathers on nonlinear
lattices \cite{pin3,Takeno1988}.
These breather solutions represent  states of excitation
that can only occur in nonlinear systems. Among other things they
allow for the observation of different properties of the materials
\cite{Lazarides2011}, which is a useful application  of breather 
oscillations.
On discrete lattices breather oscillations are usually referred to as
discrete breathers, or intrinsic localized modes. They have been  
investigated on a variety of lattices \cite{pin3,Takeno1988,Flach2008}, and,
in particular, in a one-dimensional \PT-symmetric nonlinear metamaterial 
\cite{Tsironis2013}.

The purpose of this  work is to extend these studies to
{\em two-dimensional} \PT-symmetric nonlinear metasurfaces with balanced
gain and loss.
By numerically solving the equations governing the planar array of coupled split-ring resonators we find another type of localized oscillations, which we call
hybrid breather oscillations.
In Sec. \ref{sec:sao} we set up the equations of the \PT-symmetric
two-dimensional metamaterial system and  solve these numerically. We
demonstrate that both breather and hybrid breather solutions exist in this
system. In Sec. \ref{sec:theory} we present an analytical model which helps to
explain the appearance of hybrid breather solutions. These turn out to be 
superpositions of breathers and plane waves, in such a way
that they are localized in one dimension and are extended
in the other dimension. A short summary will be given in Sec. \ref{sec:sum}.

\section{System and Numerical Results}
\label{sec:sao}

\begin{figure}
  \includegraphics[width=\columnwidth]{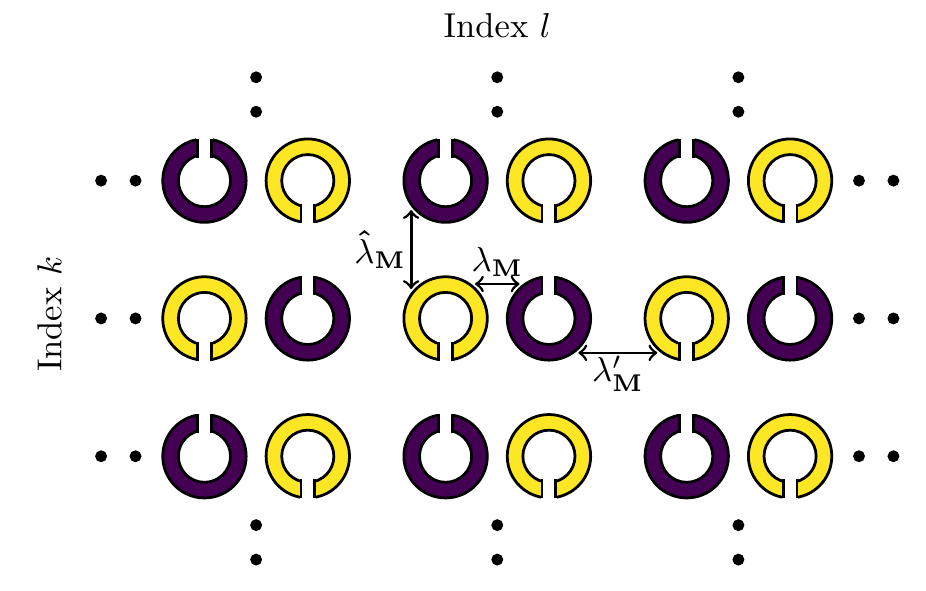}
  \caption{A nonlinear magnetically coupled metasurface with balanced gain and
    loss, described by Eq.~\eqref{eq:allg_Dimeroberflaeche}. Purple (dark) and
    yellow (bright): split-ring resonators with loss or gain, respectively.}
  \label{fig:Metaoberflaeche}
\end{figure}

We consider a two-dimensional array of dimers, each comprising two
nonlinear split-ring resonators, one with loss and the other with an equal
amount of gain (Fig.~\ref{fig:Metaoberflaeche}).
Evidently this is the two-dimensional extension of the one-dimensional chain
discussed in Ref.~\cite{Tsironis2013}. As in that reference, the split-ring
resonators are coupled magnetically through dipole-dipole forces, and are
regarded as $RLC$ circuits.
The split-ring resonators are arranged in an alternating fashion
on the two-dimensional lattice, 
in such a way that each split-ring resonator with absorption (loss)
is only surrounded by split-ring resonators with amplification (gain),
and vice versa.
Within each dimer, the split-ring resonators are coupled via $\lM$,
and the coupling between neighboring dimers is given by $\lMp$ in the
horizontal and by $\lMh$ in the vertical direction. Without loss of
generality we choose $|\lM| > |\lMp|$. 
Note that the electrical coupling between the split-ring resonators can be
neglected due to the relative orientations of the different split-ring
resonators \cite{Hesmer2007}. No external driving voltage is applied to the
metamaterial.

The equations of motion for the dynamics of the charge $q_{l,k}$ in the
capacitor at lattice site $(l,k)$  can be derived in analogy with the
one-dimensional case \cite{Tsironis2013}. The starting point is the equivalent
circuit model picture, where nearest-neighbor couplings as marked in
Fig.~\ref{fig:Metaoberflaeche} are taken into account. A detailed derivation
of the one-dimensional case can be found in Ref.\ \cite{Tsironis2006}. In
addition to this case the couplings via the constants $\hat{\lambda}_M$
are added in the two-dimensional plane. This results in the coupled differential
equations
\begin{subequations}
  \label{eq:allg_Dimeroberflaeche}
  \begin{align}
    \begin{split}
      & \lambda_M ' \ddot{q}_{2l,2k+1} 
      + \hat{\lambda}_M  \ddot{q}_{2l+1,2k} 
      + \ddot{q}_{2l+1,2k+1} \\
      + & \lambda_M \ddot{q}_{2l+2,2k+1}
      + \hat{\lambda}_M \ddot{q}_{2l+1,2k+2} 
      + q_{2l+1,2k+1} \\
      + & F(q_{2l+1,2k+1}) 
      + \gamma \dot{q}_{2l+1,2k+1} = 0\, ,
    \end{split}\\
    \begin{split}
      & \lambda_M \ddot{q}_{2l-1,2k+1} 
      + \hat{\lambda}_M  \ddot{q}_{2l,2k} 
      + \ddot{q}_{2l,2k+1} \\ 
      + & \lambda_M ' \ddot{q}_{2l+1,2k+1}
      + \hat{\lambda}_M \ddot{q}_{2l,2k+2}  
      + q_{2l,2k+1} 
      \\ 
      + & F(q_{2l,2k+1})   
      - \gamma \dot{q}_{2l,2k+1} = 0 \, ,
    \end{split}\\
    \begin{split}
      & \lambda_M ' \ddot{q}_{2l,2k} 
      + \hat{\lambda}_M \ddot{q}_{2l+1,2k-1} 
      + \ddot{q}_{2l+1,2k} \\ 
      + & \lambda_M \ddot{q}_{2l+2,2k}
      + \hat{\lambda}_M  \ddot{q}_{2l+1,2k+1}  
      + q_{2l+1,2k} \\
      + & F(q_{2l+1,2k}) 
      - \gamma \dot{q}_{2l+1,2k} = 0 \, ,
    \end{split}\\
    \begin{split}
      & \lambda_M \ddot{q}_{2l-1,2k} 
      + \hat{\lambda}_M \ddot{q}_{2l,2k-1} 
      + \ddot{q}_{2l,2k} \\
      + & \lambda_M ' \ddot{q}_{2l+1,2k}
      + \hat{\lambda}_M  \ddot{q}_{2l,2k+1}  
      + q_{2l,2k} \\ 
      + & F(q_{2l,2k})   
      + \gamma \dot{q}_{2l,2k} = 0 
    \end{split}
  \end{align}
\end{subequations}
for each site in a $2\times 2$ unit cell.
Here, $l \in \mathbb{N}$ counts the resonators in the horizontal metachains
(rows), and $k \in \mathbb{N}$ in the vertical metachains (columns). The
quantity $\gamma$ is the gain and loss parameter  for the individual split-ring
resonators, and the function $F$ describes the non-linearity. We work in
dimensionless units. The equations are normalized to the eigenfrequency of one
linear split-ring resonator. 

Amplification can be realized by an injection of energy through a tunneling
(Esaki) diode \cite{Esaki}, which features a negative ohmic resistance. 
The nonlinearity of each individual split-ring resonator can be realized by a
nonlinear dielectric which is introduced into the capacitance of each
split-ring resonator. For the numerical calculations a typical nonlinearity
for a diode is chosen, $F(q_{x,y}) = \alpha q_{x,y}^2 + \beta q_{x,y}^3$ with
$\alpha = -0.4$, $\beta = 0.08$ \cite{Lazarides2011}. As the gain and loss
can influence the stability of the nonlinear system \cite{Haag14a}
we choose a moderate, but yet significantly different from zero, value of
$\gamma = \pm 0.002$.

Note that because of the different couplings $\lM$ and $\lMh$,
the array is not fully $\PT$ symmetric in two dimensions, but only
along each row, and along each column. Thus, the parity operator can be
represented either by the spatial reflection at the center of a row \emph{or}
of a column. In both cases the time reversal operator is given by a complex
conjugation. This is an example of different $\PT$ symmetries in two dimensions,
as discussed, e.g., in Ref.~\cite{CastroAlvaredo2009} for a non-Hermitian XXZ
spin chain. The solutions of the two-dimensional split ring array have to
obey both $\PT$ symmetries to ensure balanced gain and loss.

Equations \eqref{eq:allg_Dimeroberflaeche} are solved numerically by
setting $q_{l,k} \propto \exp[i(l \kappa_l + k \kappa_k -\Omega \tau)]$
(with 4 different constants of proportionality), assuming a $1/\cosh$-type
initial charge distribution around the central row, and performing a root
search.

\begin{figure}
  \includegraphics[width=\columnwidth]{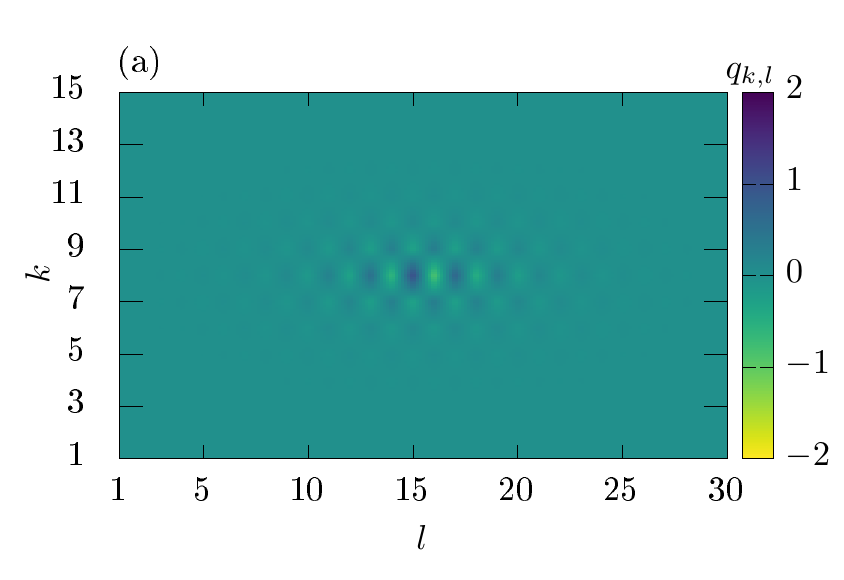}\\
  \includegraphics[width=\columnwidth]{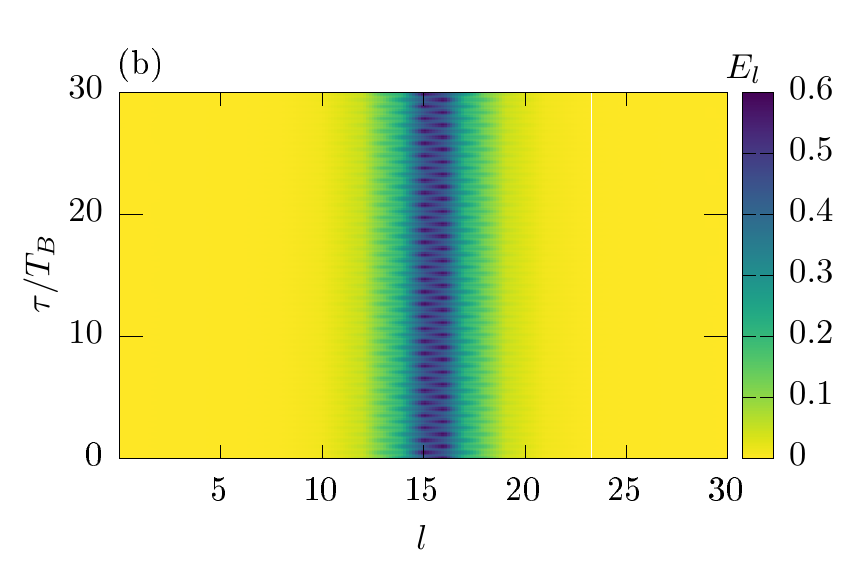}\\
  \includegraphics[width=\columnwidth]{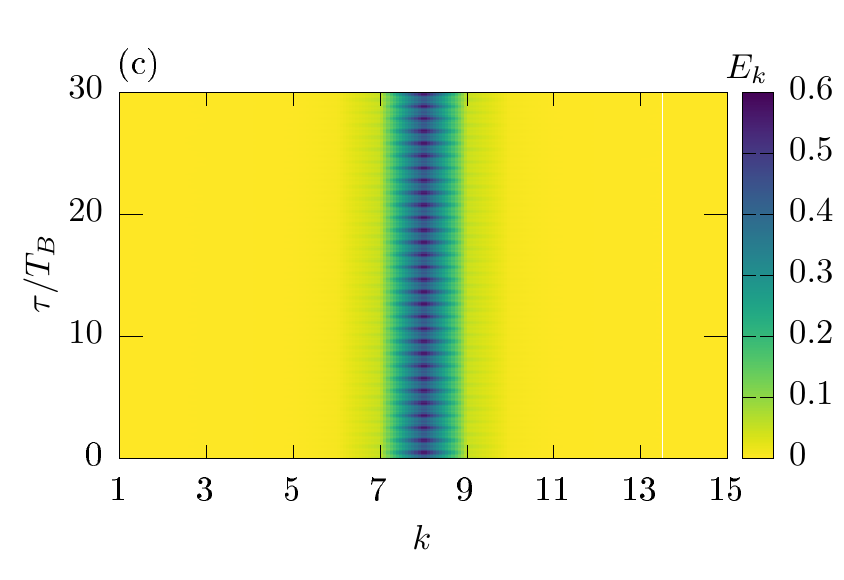}
  \caption{Breather solution with a frequency $\Omega_b = 0.8666$ for a
    system with $k = 1\dots 15$, $l = 1\dots 30$, $\lM = -0.17$, $\lMp = -0.1$,
    $\lMh = -0.015$, $\gamma = \pm 0.002$, (a) breather profile of the system,
    (b) energy of the central column, (c) energy of the central row in
    multiples of the breather period $T_B$.
    \label{fig:Breatherloesung_Metaoberflaeche_Breather_lower}}
\end{figure}

We find two-dimensional nonlinear localized, stable breather solutions.
An example with frequency $\Omega_b=0.8666$ is shown in
Fig.~\ref{fig:Breatherloesung_Metaoberflaeche_Breather_lower}.
Figure~\ref{fig:Breatherloesung_Metaoberflaeche_Breather_lower}(a) shows the
breather profile, which is strongly localized in the center of the
metamaterial and exponentially decaying towards its borders. The energy
distribution of the central column $k=8$, depicted in
Fig.~\ref{fig:Breatherloesung_Metaoberflaeche_Breather_lower}(b), and the
energy of the central row $l=15$,
Fig.~\ref{fig:Breatherloesung_Metaoberflaeche_Breather_lower}(c), clearly
exhibit the oscillating behavior of the central split-ring resonators over
30 periods of the solution. It can be seen that in both cases the energy
oscillates in a stable fashion and does not disperse along either the central
column or the central row. 

\begin{figure}
  \includegraphics[width=\columnwidth]{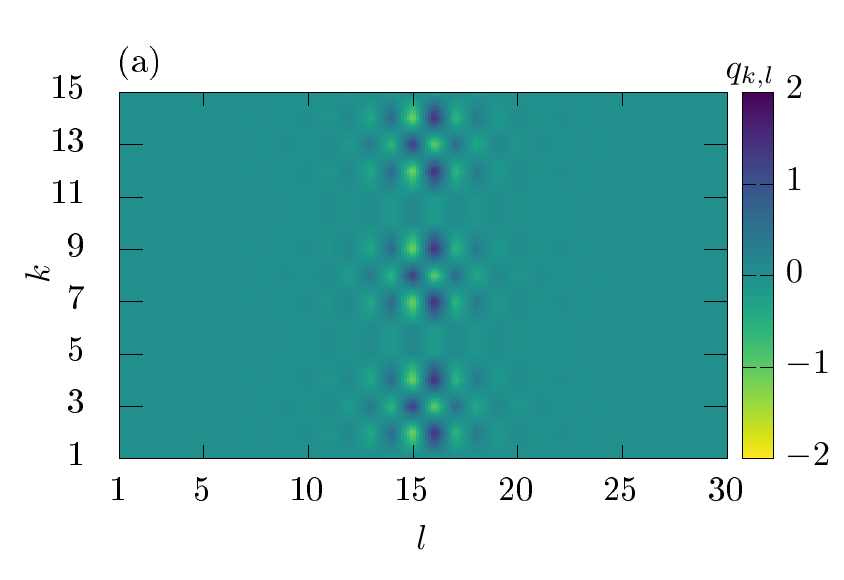}\\
  \includegraphics[width=\columnwidth]{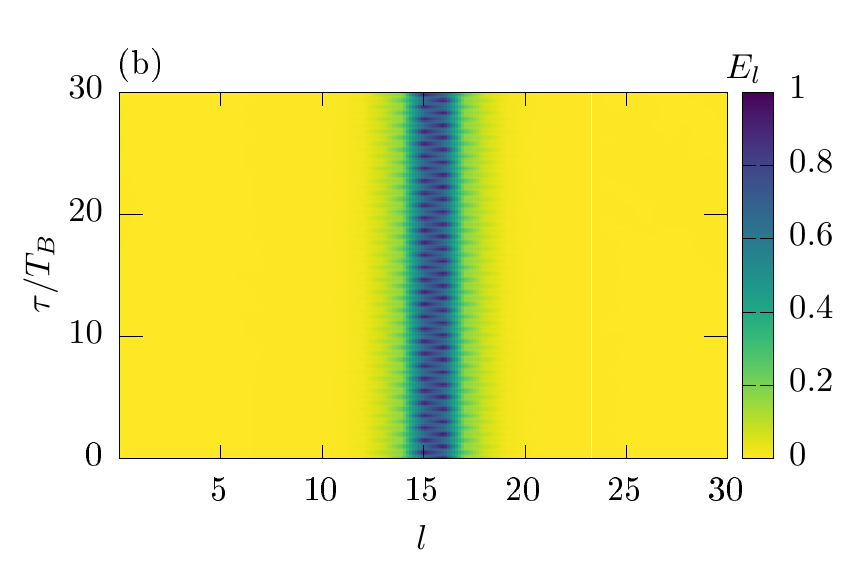}\\
  \includegraphics[width=\columnwidth]{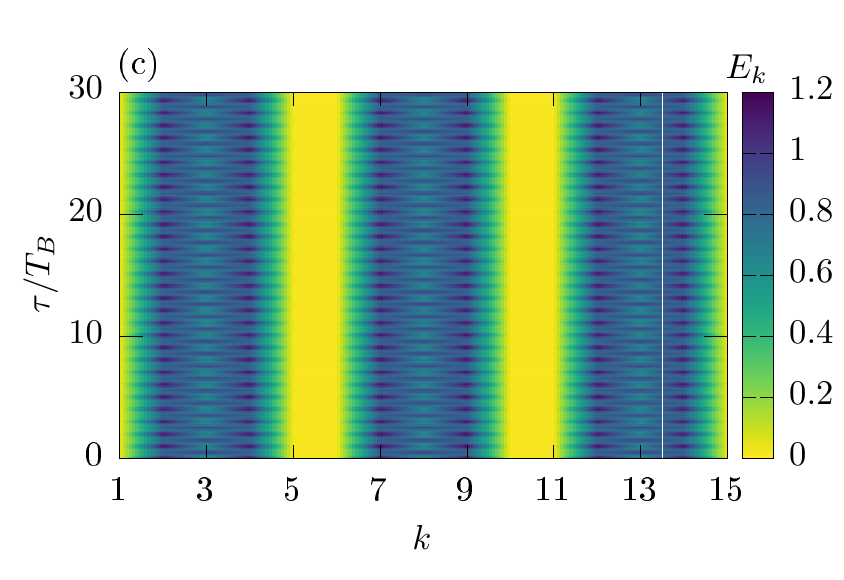}
  \caption{Hybrid breather solution, with breather part along the
    $l$-direction, (a) breather profile of the system, (b) energy of the central
    column, (c) energy of the central row in multiples of the breather period
    $T_B$ for a system with $k = 1\dots 15$, $l = 1\dots 30$, $\lM = -0.17$,
    $\lMp = -0.1$, $\lMh = -0.015$, $\gamma = \pm 0.002$.
    \label{fig:Breatherloesung_Metaoberflaeche_Mischform_row}}
\end{figure}

While numerically calculating breather solutions on this metamaterial
\eqref{eq:allg_Dimeroberflaeche}, we also detected other types of solutions such
as the one shown in
Fig.~\ref{fig:Breatherloesung_Metaoberflaeche_Mischform_row}.

Here, a sinusoidal initial charge distribution in the $k$ direction and a
localized shape around the middle of the $l$ direction was chosen. We find
again a breather type oscillation of the central column $k = 8$,
Fig.~\ref{fig:Breatherloesung_Metaoberflaeche_Mischform_row}(b), but, on the
other, several oscillations around the central row $l = 15$,
Fig.~\ref{fig:Breatherloesung_Metaoberflaeche_Mischform_row}(c).
It can be seen that in both cases the energy oscillates in a stable manner,
with no dispersion of the energy distributions. The stability of this type of
solution corresponds to that of the two-dimensional breather solution shown
in Fig.~\ref{fig:Breatherloesung_Metaoberflaeche_Breather_lower}.

These oscillations have strong resemblance to a breather solution, however,
it is definitely not a two-dimensional breather. To understand the behavior of
this special type of breather oscillation, which we call hybrid breather, we
will consider an analytical model which starts from the superposition of a
breather solution and a plane wave solution in the next section.

\section{Analytical Results}
\label{sec:theory}
The breather spectrum of a system can be calculated analytically with the usual
approximation of dividing the breather oscillation into two major parts
\cite{pin3}, the nonlinear central part, also called the breather core, and the
linearized part at the borders, called breather tails. At the borders the
linearized coupled differential equations of the system are solved with the
breather tails ansatz. This separation approximation for the analytical
breather solution can be chosen because we are looking for spatially decaying
solutions on a lattice, which have the characteristic that after a certain
lattice site the amplitude of the onsite oscillations is small enough to
neglect the nonlinear interactions.

To obtain the spectrum of a hybrid breather solution an ansatz 
\begin{align}
  \begin{split}
    q_{2x,2y+1} = & A e^{ 2x \kappa_X + i ((2y+1) \kappa_Y - \Omega \tau)} \, , \\
    q_{2x+1,2y+1}  = & B e^{ (2x+1) \kappa_X + i((2y+1) \kappa_Y - \Omega \tau)}\, ,\\
    q_{2x,2y}  = & C e^{ 2x \kappa_X + i(2y \kappa_Y - \Omega \tau)} \, ,\\ 
    q_{2x+1,2y} = & D e^{ (2x+1) \kappa_X + i(2y \kappa_Y - \Omega \tau)} \\
  \end{split}
  \label{eq:Breatherloesung_Metaoberflaeche_Mischform_Ansatz}
\end{align} 
with the normalized wave vectors $\kappa_{X,Y} \in \mathbb{R}$, is inserted
into the linearized coupled differential equations of the system
\eqref{eq:allg_Dimeroberflaeche}, which is similar to a main breather
frequency ansatz in one dimension denoted by $X$ and a plane wave ansatz in
the other dimension called $Y$.
The arbitrary coordinates $X$ and $Y$ for the hybrid breather solution are
chosen to show the independence of the existence of hybrid breather solutions
from the direction of the breather or the plane wave part.

By requiring nontrivial solutions for the resulting stationary problem one
obtains
\begin{align}
  \Omega_{\gamma;\pm} = \Omega^4 \left( \sqrt{ \lambda_{\mu}} \pm \hat{\mu}_{k}
  \right)^2 \, ,
\end{align}
where $\lambda_{\mu} = (\lM - \lMp)^2 + \mul \mulp$, $\Omega_{\gamma}
= (1-\Omega^2)^2 + \gamma^2 \Omega^2$. Solving the equation with respect to
$\Omega$ yields four frequencies, and because of the $\PT$ symmetry just
positive frequency solutions are considered, 
\begin{equation}
  \Omega_{b;\vec{\kappa};\pm,\pm} = \sqrt{\frac{1 - \frac{\gamma^2}{2}
      \pm \sqrt{\frac{\gamma^4}{4} - \gamma^2 + (\sqrt{\lambda_{\mu}}
        \pm \hat{\mu}_k)^2 }}{[1 - (\sqrt{\lambda_{\mu}} \pm \hat{\mu}_k)^2]}}
  \, .
  \label{eq:Breatherloesung_Metaoberflaeche_Breather_Omega}
\end{equation}
For a hybrid breather solution, with a breather shape along the dimer chains
in $l$-direction, the substituted variables are
\begin{align}
  \begin{split}
    \mu_{l} = & 2 \lambda_M \cosh(\kappa_l) \, , \quad \mu_{l}'
    = 2 \lambda_M' \cosh(\kappa_l) \, , \\ 
    \hat{\mu}_{k} = & 2  \hat{\lambda}_M \cos(\kappa_k) \, ,
  \end{split}
                      \label{eq:l}
\end{align}
and with a breather shape orthogonal to the dimer chains in $k$-direction
\begin{align}
  \begin{split}
    \mu_{l} = & 2 \lambda_M \cos(\kappa_l) \, , \quad \mu_{l}'
    = 2 \lambda_M' \cos(\kappa_l) \, , \\ 
    \hat{\mu}_{k} = & 2  \hat{\lambda}_M \cosh(\kappa_k) \, .
  \end{split}
                      \label{eq:k}
\end{align}

\begin{figure*}
  \centering
  \begin{tabular}{@{}c@{}c@{}}
  \includegraphics[width=0.45\textwidth]{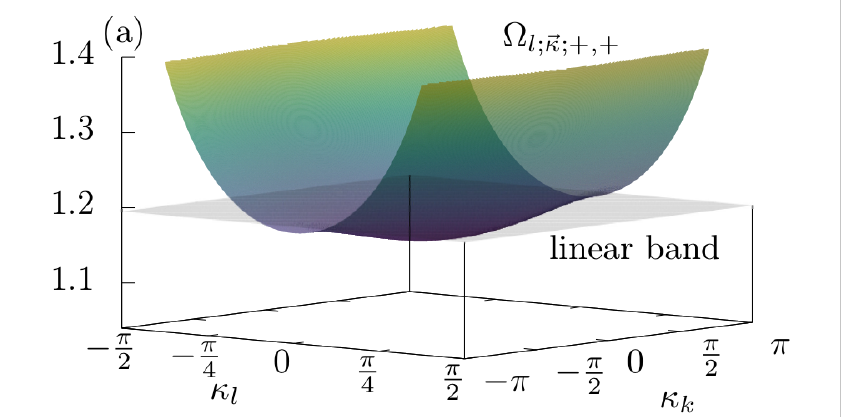} &
  \includegraphics[width=0.45\textwidth]{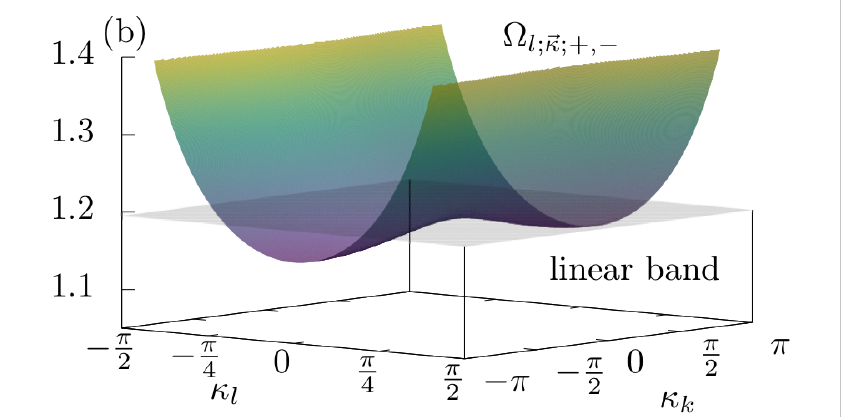} \\
  \includegraphics[width=0.45\textwidth]{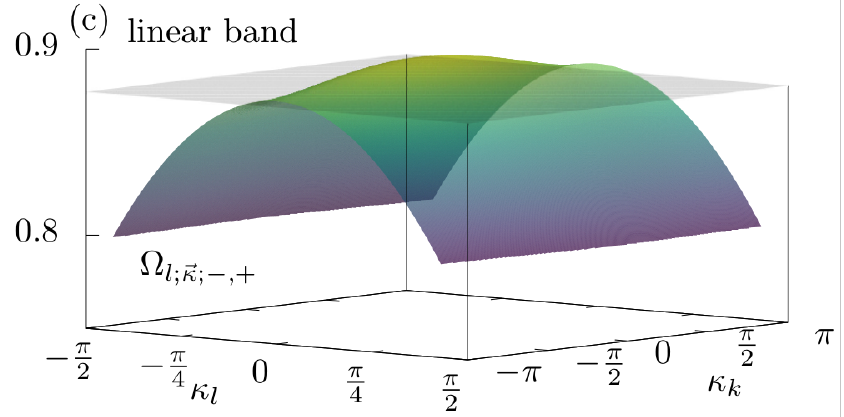} &
  \includegraphics[width=0.45\textwidth]{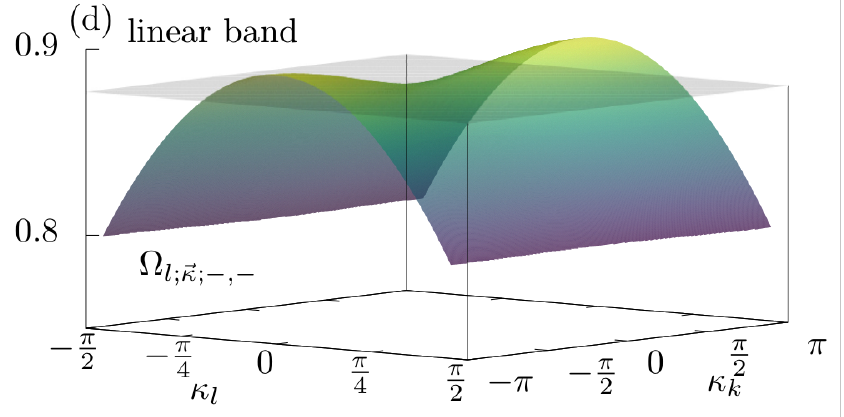} \\
  \includegraphics[width=0.45\textwidth]{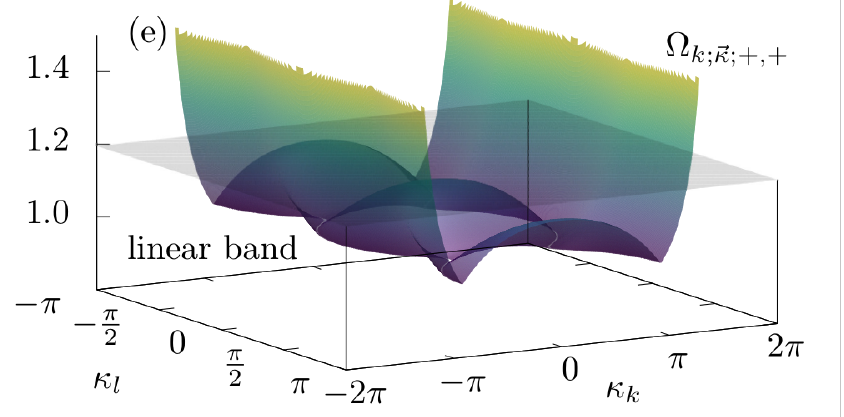} &
  \includegraphics[width=0.45\textwidth]{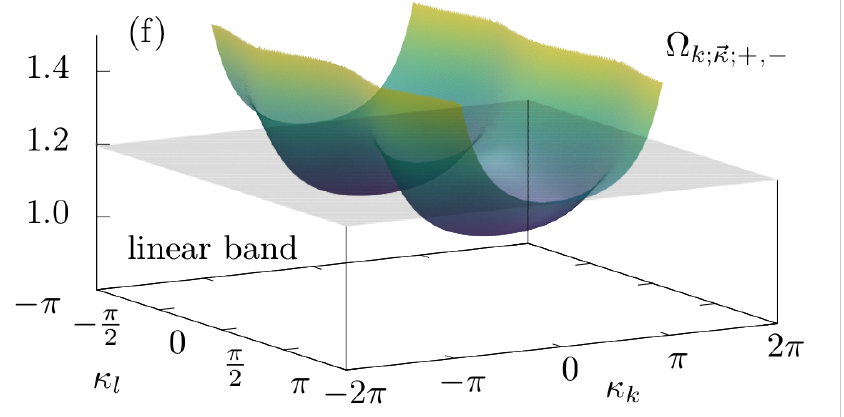} \\
  \includegraphics[width=0.45\textwidth]{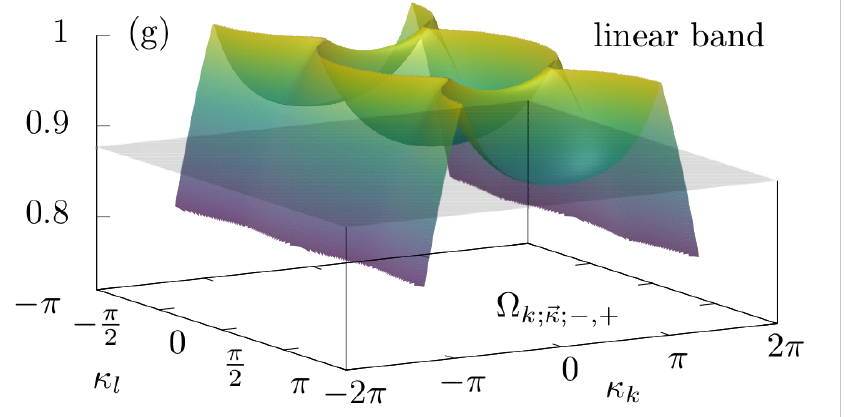} &
  \includegraphics[width=0.45\textwidth]{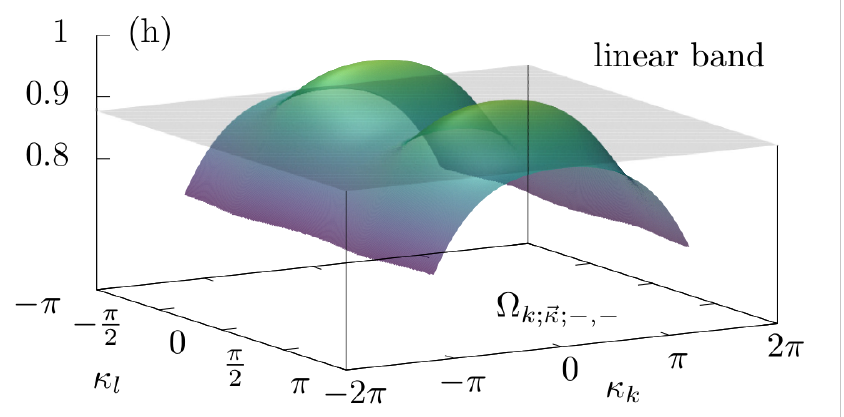}
  \end{tabular}
  \caption{Analytically approximated hybrid breather solutions obtained
    with Eq.~\eqref{eq:Breatherloesung_Metaoberflaeche_Breather_Omega}
    for coupling constants $\lambda_M = -0.17$, $\lambda_M' = -0.1$ and
    $\hat{\lambda}_M = -0.015$, (a) $\Omega_{l;\vec{\kappa};+,+}$,
    (b) $\Omega_{l;\vec{\kappa};+,-}$, (c) $\Omega_{l;\vec{\kappa};-,+}$,
    (d) $\Omega_{l;\vec{\kappa};-,-}$ with a breather shape along the
    dimer chains in $l$-direction, (e) $\Omega_{k;\vec{\kappa};+,+}$,
    (f) $\Omega_{k;\vec{\kappa};+,-}$, (g) $\Omega_{k;\vec{\kappa};-,+}$,
    (h) $\Omega_{k;\vec{\kappa};-,-}$ with a breather shape orthogonal to
    the dimer chains in $k$-direction. The gray planes indicate the
    frequency of the lower and upper edges of the allowed linear
    frequency band. Explicitly, the breather condition reads
    $\Omega < 0.8771$ or $\Omega > 1.1952$.
    \label{fig:Breatherloesung_Metaoberflaeche_Mischform_Spektrum}}
\end{figure*}

These four analytically calculated solutions are displayed in
Fig.~\ref{fig:Breatherloesung_Metaoberflaeche_Mischform_Spektrum}
for the same parameters as used in  the numerically calculated solution
(Fig.~\ref{fig:Breatherloesung_Metaoberflaeche_Mischform_row}),  and for both
cases of the substituted variables given in Eqs.~\eqref{eq:l} and
\eqref{eq:k}. In
Figs.~\ref{fig:Breatherloesung_Metaoberflaeche_Mischform_Spektrum}(a)-(d) the
four solutions are displayed for a breather shape along the dimer chains in
$l$-direction. It can be seen that the solutions $\Omega_{b;\vec{\kappa};+,+}$
in Fig.~\ref{fig:Breatherloesung_Metaoberflaeche_Mischform_Spektrum}(a) and
$\Omega_{b;\vec{\kappa};+,-}$ in
Fig.~\ref{fig:Breatherloesung_Metaoberflaeche_Mischform_Spektrum}(b) as well as
$\Omega_{b;\vec{\kappa};-,+}$ in
Fig.~\ref{fig:Breatherloesung_Metaoberflaeche_Mischform_Spektrum}(c) and 
$\Omega_{b;\vec{\kappa};-,-}$ in
Fig.~\ref{fig:Breatherloesung_Metaoberflaeche_Mischform_Spektrum}(d) are
similar to each other and can be transformed into one another by a shift of
$\Delta \kappa_k = \pi$. In
Figs.~\ref{fig:Breatherloesung_Metaoberflaeche_Mischform_Spektrum}(e)-(h) the
four solutions are displayed for a breather shape orthogonal to the dimer
chains in $k$-direction. With
Eqs.~\eqref{eq:Breatherloesung_Metaoberflaeche_Breather_Omega} and
\eqref{eq:k} it can be seen that the solutions $\Omega_{b;\vec{\kappa};+,+}$ in
Fig.~\ref{fig:Breatherloesung_Metaoberflaeche_Mischform_Spektrum}(e) and
$\Omega_{b;\vec{\kappa};+,-}$ in
Fig.~\ref{fig:Breatherloesung_Metaoberflaeche_Mischform_Spektrum}(f) as well as
$\Omega_{b;\vec{\kappa};-,+}$ in
Fig.~\ref{fig:Breatherloesung_Metaoberflaeche_Mischform_Spektrum}(g) and 
$\Omega_{b;\vec{\kappa};-,-}$ in
Fig.~\ref{fig:Breatherloesung_Metaoberflaeche_Mischform_Spektrum}(h) are the
same for $\kappa_k = 0$. A closer look at these pairs of solutions reveals
that for small values of $\kappa_k$ the solutions differ from each other but
merge again into each other for higher values of $\kappa_k$. 

To obtain a true hybrid breather solution these four solutions with the
pseudo wave vector  $\kappa = \left( \kappa_l, \, \kappa_k \right)^T$ have
also to fulfill the breather condition that their frequencies must not lie in
the (allowed) frequency band of plane wave solutions
\cite{Flach2008}
\begin{equation}
  \Omega_b \neq \Omega_{\text{plane wave}} \, ,
  \label{eq:breather_condition}
\end{equation}
which is obtained by solving the linearized system equations. For the
parameters considered here the lower edge of the band is found to be
$\Omega = 0.8771$ and the upper edge $\Omega = 1.1952$. The breather condition
\eqref{eq:breather_condition} is requisite to guarantee the existence of a
stable breather core  of the hybrid breather oscillation.
Out of this condition not every pseudo wave vector of all four analytically
calculated solutions belongs to a frequency in the allowed domain.
It can be seen in
Fig.~\ref{fig:Breatherloesung_Metaoberflaeche_Mischform_Spektrum} that
for the solutions $\Omega_{b;\vec{\kappa};+,+}$ and $\Omega_{b;\vec{\kappa};+,-}$ 
only hybrid breather solutions with a frequency above the upper edge of
the allowed frequency band ($\Omega = 1.1952$) are found.
Analogously this can be applied to $\Omega_{b;\vec{\kappa};-,+}$ and
$\Omega_{b;\vec{\kappa};-,-}$, where the pseudo vector needs to belong to a
frequency under the lower edge of the allowed frequency band
($\Omega = 0.8771$).

For the analytical breather solution with a breather shape along the
$k$-direction two favorable solutions $\Omega_{b;\vec{\kappa};+,-}$ and 
$\Omega_{b;\vec{\kappa};-,-}$ exist. They have a much smaller range of pseudo
wave vectors which belong to frequencies inside the linear band structure.
Thus, these solutions can describe true hybrid breather solutions for more
pseudo wave vectors than the other two analytical solutions. On the other hand
the analytical breather solutions with a breather shape along the $l$-direction
do not have any favorable solutions, because each solution has the same amount
of pseudo wave vectors belonging to frequencies inside the linear band
structure.

It is interesting to note that if the analytical $\Omega_{b;\vec{\kappa};+,+}$
hybrid breather solution fulfills the breather condition
for a certain pseudo wave vector $\kappa$, then also
$\Omega_{b;\vec{\kappa};-,+}$ fulfills the condition for the same pseudo
wave vector, and vice versa. The same can be observed for
$\Omega_{b;\vec{\kappa};+,-}$ and $\Omega_{b;\vec{\kappa};-,-}$,
which is shown in
Fig.~\ref{fig:Breatherloesung_Metaoberflaeche_Mischform_Spektrum}.

The analytical model was calculated using the linearization approximation
for the breather tails. It covers all solutions that contain exponentially
dropping tails. This is independent from the nonlinearity chosen since
we only look at those parts of the tails which have an amplitude small
enough for the nonlinearities to be insignificant. Therefore this model
describes every possible hybrid breather excitation caused by some
nonlinearity in the breather core. However, the numerically observable
hybrid breather oscillations depend strongly on the nonlinearity chosen
for the system and the shape of the breather core. Analogously
to the normal breather oscillations the hybrid breather oscillations have
to obey \PT symmetry to obtain long lived oscillations. The nonlinearities
chosen in this paper represent a typical diode \cite{Tsironis2013}.

\begin{figure}
  \includegraphics[width=\columnwidth]{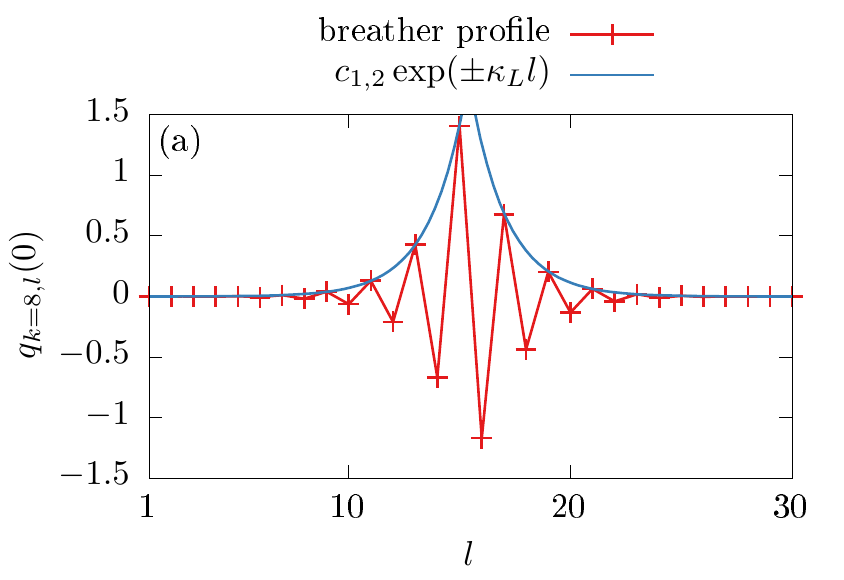}\\
  \includegraphics[width=\columnwidth]{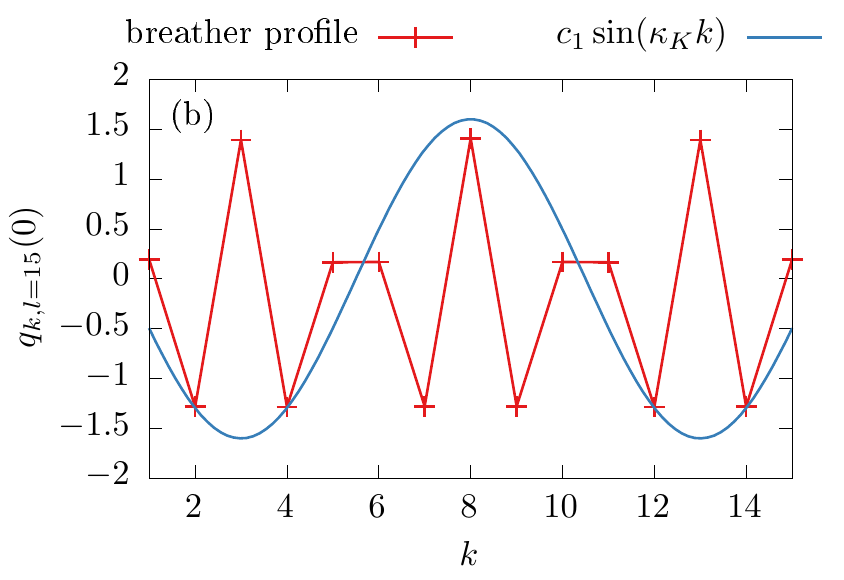}
  \caption{Comparison between the breather profile of the numerically
    calculated system
    (Fig.~\ref{fig:Breatherloesung_Metaoberflaeche_Mischform_row}) and the
    analytical model
    (Fig.~\ref{fig:Breatherloesung_Metaoberflaeche_Mischform_Spektrum}) with
    the same coupling constants for (a) the central column (8th) and (b) the
    central row (15th) of the system.
    \label{fig:Breatherloesung_Metaoberflaeche_Mischform_row_compare}}
\end{figure}

The results of the analytical theory can now be compared with the numerically
observed hybrid breather oscillation, shown in
Fig.~\ref{fig:Breatherloesung_Metaoberflaeche_Mischform_row}, with a standing
plane wave form along the $k$-direction.
In Fig.~\ref{fig:Breatherloesung_Metaoberflaeche_Mischform_row_compare}(a)
we compare the breather profile along the central column: the breather part
for the numerically calculated hybrid breather solution (red lines with
crosses), and the analytical exponential functions (blue solid lines). 
As can be seen in the diagram the analytical model describes the numerical
values perfectly. Through this a value of $\kappa_l = 0.6$ was determined for
the numerical solution. In
Fig.~\ref{fig:Breatherloesung_Metaoberflaeche_Mischform_row_compare}(b) we
compare the breather profile along the central row for the numerically
calculated hybrid breather solution, the standing plane wave part (red lines
with crosses) with an analytical sine function (blue solid lines). As can be
seen in the diagram the sine function generally describes the numerical values
quite well. Discrepancies between the numerical and the analytical solution
occur because we are numerically calculating a nonlinear system with charges
$|q_{k,l}(\tau/T_B)| > 1$ in which case also the nonlinearity of the split-ring
resonators would have to be taken into account. This nonlinearity of the
split-ring resonators was neglected in the analytical calculation due to the
linearization approximation of the breather tails. From the comparison a value
of  $\kappa_k = \pi /5$ is deduced for the numerical solution. Inserting the
numerical parameters into
Eq.~\eqref{eq:Breatherloesung_Metaoberflaeche_Breather_Omega}
yields the four frequency values $\Omega_{+,+} = 1.1890$, $\Omega_{+,-}
= 1.2321$, $\Omega_{-,+} = 0.8795$, $\Omega_{-,-} = 0.8635$. By comparing these
values with the numerical ones it can be seen that the frequency $\Omega_{-,-}
= 0.8635$ and the numerical ones are equivalent. For this reason the   
analytical approximation describes the hybrid breather solutions quite well 
for plane wave parts slightly in the nonlinear regime.

In the analytical approximation also hybrid breather solutions with a plane
wave part along the $l$-direction and hybrid breather solutions with a plane
wave part along the $k$-direction and a frequency above the linear band
structure are predicted. However such hybrid breather solutions do not appear
in the numerical calculation.

An explanation for this can be found in the analytical model. To obtain a
stable breather oscillation a nonlinear breather core is needed which
oscillates with the breather frequency and has an according shape to induce
the oscillation in the linearized breather tails. The formulated analytical
model only supports nonlinear breather cores which are similar to linear plane
waves, because the nonlinear breather core excites an oscillation outside of
the linear band structure which then will spatially decay in the linearized
breather tails. This leads to the fact that the described 
analytical model can only be used for not too large nonlinear hybrid
breather cores. As a consequence we cannot find hybrid breather solutions with
a plane wave part along the $l$-direction and a frequency above the linear
band structure in our numerical calculations. One way to find the other
solutions would be to extend our analytical model to nonlinear plane waves.
Another way would be the use of a metamaterial with a different nonlinearity
and other breather cores.

The breather cores that can be found strongly depend on the nonlinearity of the 
metamaterial. By contrast the breather tails are totally independent of the
nonlinearity of the metamaterial. Therefore it is possible that similar
metamaterials with slightly differing nonlinearities can exhibit quite different
breather and hybrid breather solutions. This is also the reason why we
evaluated all hybrid breather  solutions $\Omega_{b;\vec{\kappa};\pm,\pm}$,
because for every possible hybrid breather solution there could exist
a nonlinear metamaterial with an appropriate nonlinearity with which it is
possible to excite this hybrid breather oscillation.

\section{Summary}
\label{sec:sum}
The main result of our work is that for \PT-symmetric nonlinear metamaterials
with balanced gain and loss in dimensions $d > 1$ mixed types of breather
oscillations can exist, in addition to usual breathers.

We have shown this by numerically  solving the equations of motion describing
the dynamics of the charges in the individual split-ring capacitors of a
\PT-symmetric nonlinear metasurface.  These, as we call them,
`hybrid breather' solutions could be explained by an analytical model, in which
the breather was divided into a nonlinear central part and a linearized outer
part, and by allowing a linear plane wave shape in one direction.

\end{document}